\documentclass[prd,showpacs,nofootinbib,twocolumn,amsmath,amsfonts,amssymb]{revtex4} 

\usepackage{amsmath}
\usepackage{amssymb}
\usepackage{epsfig}

\usepackage{graphicx}
\usepackage{stmaryrd}
\usepackage{bm}
\usepackage{color}

\newcommand{\AddrAHEP}{%
 Cluster of Excellence, Origin and Structure of the Universe, Technische Universit\"{a}t M\"{u}nchen,\\
 Boltzmannstra\ss e 2, D-85748 Garching, Germany}

\begin{document}

\title{An estimate of $\theta_{14}$ independent of the reactor antineutrino flux determinations}

\author{Antonio Palazzo}
\affiliation{\AddrAHEP}

\date{\today}

\begin{abstract}

  In a previous paper [Phys.\ Rev.\ {\bf D} 83, 113013 (2011)] we have shown 
  that the solar sector data (solar and KamLAND) are sensitive to the parameter $\theta_{14}$, encoding
   the admixture of the electron neutrino with a fourth (essentially) sterile mass eigenstate. 
   In that work we evidenced that such data prefer a non-zero value of
   $\theta_{14}$ and that such a preference is completely degenerate with that  of non-zero 
   $\theta_{13}$.  In this report we show how the  evidence of  $\theta_{13}>0$, recently 
   emerged from global neutrino data analyses, lifts such a degeneracy and disfavors
   the case of sterile neutrino mixing. By excluding from our analysis the total rate information coming
  from the reactor experiments we untie our results from any assumption on their flux normalization. 
   In this way, we establish the robust upper bound $ \sin^2 \theta_{14} < 0.04$ at the 90\%  C.L.  

\end{abstract}


\maketitle

In a recent paper~\cite{Palazzo:2011rj} we have introduced  the theoretical framework 
needed to describe solar neutrino oscillations within the so-called
$3+ s$ schemes endowed with $s$ new sterile neutrinos (see also~\cite{Giunti:2009xz}). In the same work,
we have considered the constraints attainable within such schemes from the ``solar sector'' (solar and KamLAND data)  
showing that this dataset, while preferring a non-null
admixture of the electron-neutrino with mass eigenstates ``far'' from 
the solar ($\nu_1, \nu_2 $) doublet,  is  currently unable to distinguish  if such a
 mixing is realized with the third standard 
mass eigenstate $\nu_3$ or with new ones ($\nu_{3+1}, ..., \nu_{3+s}$). In the
simplest $3+1$ framework, this ambiguity translates into a degeneracy of the
estimates of the standard mixing angle $\theta_{13}$ and the new angle $\theta_{14}$
(see~\cite{Palazzo:2011rj}  for the details of the parameterization of the lepton mixing matrix).

After publication of~\cite{Palazzo:2011rj} new data have been released  that are relevant
to the analysis therein performed. In particular, the long-baseline (LBL) accelerator experiments 
Tokai-to-Kamioka (T2K)~\cite{Abe:2011sj} and the Main Injector Neutrino Oscillation Search (MINOS)~\cite{Adamson:2011qu} 
have both evidenced a phenomenon of $\nu_\mu \to \nu_e$ conversion. Moreover,
the reactor experiment Double-CHOOZ (D-CHOOZ)~\cite{Abe:2011fz}, currently operating 
only with the far detector, has found an indication of $\nu_e \to \nu_e$ disappearance. 
These findings, if interpreted within the standard 3-flavor framework, point towards a
non-zero value of $\theta_{13}$, in line with the first indications arising from 
global neutrino data analysis~\cite{Fogli:2008jx}
(see also~\cite{Schwetz:2008er,GonzalezGarcia:2010er}).
In fact, with the inclusion of the new crucial piece 
of information, an updated global neutrino data analysis~\cite{Fogli:2011qn}
(see also~\cite{Schwetz:2011zk, Machado:2011ar}) provides%
\footnote{The analyses in~\cite{Fogli:2011qn, Schwetz:2011zk} do not incorporate the D-CHOOZ result,
whose inclusion would further reinforce the evidence of non-zero
$\theta_{13}$ therein established. A preference for $\theta_{13}>0$ around the
3$\sigma$ level has been shown also in the analysis performed in~\cite{Machado:2011ar}, 
which includes D-CHOOZ (together with MINOS and T2K), but not  the solar and atmospheric data.} 
evidence of $\theta_{13} >0$ at more than 3$\sigma$.

This new circumstance prompts us to improve the analysis performed in~\cite{Palazzo:2011rj},
in order to determine how it is affected by the new  critical experimental information. 
Substantial changes with respect to the results presented
in~\cite{Palazzo:2011rj} are expected. In fact, due to the strong anti-correlation 
existing among the two mixing angles $\theta_{13}$ and $\theta_{14}$, the  
clear preference now emerged for a non-zero value of one of the two parameters ($\theta_{13}$)
should drastically reduce the likelihood of the other one ($\theta_{14}$)
to be different from zero. Quantifying such a qualitative expectation appears
particularly urgent in view of the numerous ongoing projects of  new experimental setups 
aimed at testing potential oscillations into sterile neutrinos (see for 
example~\cite{Formaggio:2011jt, Cribier:2011fv, Dwyer:2011xs, Novikov:2011gp,Egorov,Gorbachev,Ianni,Lasserre,Bowden}).

The new landscape brings us to adopt a more conservative approach with respect to
that espoused in~\cite{Palazzo:2011rj}, as here our prime aim is to establish
a robust estimate of  $\theta_{14}$ independent of any assumption on the determinations
of the reactor antineutrino fluxes.  Indeed, their recent recalculations~\cite{Mueller:2011nm, Huber:2011wv},
indicating an upward shift of about $3\%$ with respect to previous estimates, have
not only refueled the interest around sterile neutrinos, but have also 
engendered an intense debate around possible systematic uncertainties,
being common opinion that these may not be entirely under control.
Having this issue in mind, we will treat  the reactor data in a special way, 
minimizing the impact of the systematic uncertainties affecting the
antineutrino fluxes. More specifically, in both the short-baseline 
reactor experiments (CHOOZ and D-CHOOZ) and the long-baseline ones (KamLAND),
we will ignore the (flux dependent) {\em total rate} information, 
considering only that one provided by the energy {\em spectral shape}.

This stratagem, although slightly limiting the constraining
power of the analysis, will render its results particularly 
robust. In fact, as discussed in~\cite{Palazzo:2011rj}, the KamLAND analysis is quite sensitive
to the reactor flux normalization. In particular, the indication in favor of
non-zero $\theta_{13}$ (or $\theta_{14}$) arising from the solar sector 
fluctuates between $1.3 \sigma$ and $1.8 \sigma$,
adopting, respectively, the old or the new (higher) fluxes~\cite{Palazzo:2011rj}. 
As a rule of thumb, we have verified that  an upward (downward) 1$\%$ shift of
the reactor fluxes corresponds to a $0.15 \sigma$ increase (decrease) 
in the statistical significance of the preference for a non-zero electron
neutrino mixing with $\nu_3$ (or $\nu_4$).
By removing the KamLAND total rate information from the analysis, 
we eliminate any dependency on the reactor flux normalization.
In practice, with this procedure,  the mixing angles $\theta_{13}$  and 
$\theta_{14}$ (and to a large extent also the ``solar'' mixing  angle $\theta_{12}$)
are basically constrained by the solar data augmented%
\footnote{The solar data alone, without the ``external'' information on  $\Delta m^2_{sol}$ provided 
by the KamLAND  spectral shape, would have a reduced sensitivity to all mixing angles.
On the other hand, the KamLAND spectral shape provides little information on these last
ones.}
 by the knowledge of
the solar squared-mass difference  $\Delta m^2_{sol}$, whose
high-precision determination is preserved by retaining the KamLAND spectral shape information.

Analogous considerations apply to the CHOOZ and D-CHOOZ experiments. Also in this case 
more (less) disappearance, and thus a preference of larger (smaller) values
of $\theta_{13}$ or $\theta_{14}$, is driven by higher (lower) reactor fluxes.
Differently from KamLAND,  however, the spectral information does not give 
any information on the relevant (atmospheric) 
mass-splitting $\Delta m^2_{atm}$,  this being independently determined by the 
LBL  $\nu_\mu \to \nu_\mu$ disappearance searches performed at accelerators.
 It should be stressed that, in principle, the CHOOZ and D-CHOOZ spectral information could distinguish 
between the $\nu_3$-driven (distorted) and $\nu_4$-driven (undistorted%
\footnote{It must be stressed that at the far detector (the only one currently operational at the D-CHOOZ site)
the oscillations driven by the new mass-mixing parameters ($\Delta m^2_{new}, \theta_{14}$) get completely
averaged if $\Delta m^2_{new} \gtrsim  0.1~\mathrm{eV}^2$ (see~\cite{deGouvea:2008qk,Apollonio:2002gd}). 
Therefore, in the region of the parameter space of current interest (confined to values of  
$\Delta m^2_{new} \sim 1~ \mathrm{eV}^2$), we can safely assume that the ($\Delta m^2_{new}, \theta_{14}$)-induced
oscillations are completely averaged with a consequent undistorted energy spectrum. Of course, the situation would 
be different at a detector located near to  the reactor core (not operational at present), where non-negligible 
($\Delta m^2_{new}, \theta_{14}$)-induced spectral distortions are expected (see the discussion in~\cite{deGouvea:2008qk}). 
Finally, we remark that in the solar sector the new oscillations get averaged provided that 
$\Delta m^2_{new} \gg \Delta m^2_{sol}$, as we have shown in~\cite{Palazzo:2011rj}.})
 oscillated spectra, but its impact is negligible in practice since the expected distortions 
are very small (see the ``Analysis C'' in~\cite{Apollonio:2002gd}). 
Indeed, the observation of such spectral distortions
will be a challenge even for the next-generation of reactor experiments equipped 
with  near detectors~\cite{reactor}. The achievement of this goal appears now even more important in 
light of the opportunity of testing and distinguishing standard and non-standard physics.

\begin{figure}[t!]
\vspace*{-4.3cm}
\hspace*{-0.10cm}
\includegraphics[width=13.0 cm]{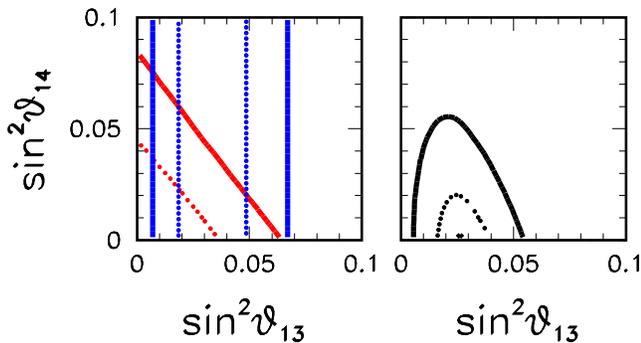}
\vspace*{-4.2cm}
\caption{Left panel: regions allowed after marginalization of the solar  ($\Delta m^2_{sol}, \theta_{12}$) 
and atmospheric ($\Delta m^2_{atm}, \theta_{23}$) mass-mixing parameters by  
the solar sector data (diagonal bands) and LBL accelerator data (vertical bands). 
Right panel: regions allowed by their combination.
The contours refer to $\Delta \chi^2 =1$ (dotted  line) 
and $\Delta \chi^2 = 4$ (solid line).
\label{fig2}}
\end{figure}  

Concerning the data sensitive to  $\Delta m^2_{sol}$ our analysis includes all the relevant solar
and KamLAND data as described in detail in~\cite{Palazzo:2011rj},  but here  the KamLAND absolute 
normalization is treated as a free parameter. As in [1] we have made the assumption that the additional
mixing angles involving sterile neutrinos are null ($\theta_{24} = \theta_{34} = 0$).%
\footnote{The assumption $\theta_{24} = \theta_{34} = 0$, implying in our parameterization (see~\cite{Palazzo:2011rj})
$U_{\mu 4}= 0$, is justified by the negative results of the short-distance disappearance searches performed in the $\nu_\mu \to \nu_\mu$ 
channel~\cite{Dydak:1983zq, Mahn:2011ea}, by the atmospheric data analyses~\cite{Maltoni:2007zf},  and by the neutral current interaction searches performed by MINOS~\cite{Adamson:2011ku}. These last ones provide the stringent upper bound $\theta_{24}<7^{\circ}$ at 
the 90\% C.L.~\cite{Adamson:2011ku}.  For such small values  the 4$\nu$-oscillation effects induced in LBL experiments, being (doubly) suppressed by the product $|U_{e4}|| U_{\mu4}|$, would have a negligible impact in our analysis. In passing, we notice that it is for the same reason that the excess of the electron-like events observed in T2K  and MINOS is not imputable to oscillations into sterile states.}




The regions allowed by the combined solar and KamLAND data represented by the diagonal bands
in the left panel of Fig.~1 show no preference for non-zero mixing.  
This behavior, which is slightly different respect to that 
observed in~\cite{Palazzo:2011rj} (where we found a weak preference for non-zero mixing), 
 can be traced to the following three  factors:
(I) The solar data taken alone give $\theta_{13} = \theta_{14} = 0$ as their best fit point;%
\footnote{This feature has been also reported in other analyses~\cite{Schwetz:2008er,  GonzalezGarcia:2010er} 
for what concern $\theta_{13}$.}
(II) The KamLAND spectral shape taken alone does not show any preference
for non-zero $\theta_{13}$ or $\theta_{14}$;%
\footnote{Within a three-flavor framework the same behavior has been observed also in~\cite{Schwetz:2011qt}.}
(III) The well-known interplay of KamLAND and solar 
data in pushing the $\theta_{13}$ ($\theta_{14}$) estimate upwards 
(see~\cite{Palazzo:2011rj,Fogli:2008jx,Balantekin:2008zm, Schwetz:2008er,  GonzalezGarcia:2010er}), 
so as to reduce the mismatch existing at $\theta_{13} = \theta_{14} =0$ among their (slightly different)
determinations of the solar mixing angle  $\theta_{12}$,
is now less effective since the KamLAND spectral shape has reduced sensitivity to this last parameter.

Concerning the data sensitive to $\Delta m^2_{atm}$, we incorporate the LBL 
accelerator results as in~\cite{Fogli:2011qn}, accounting for  the $\nu_\mu \to \nu_\mu$ 
disappearance searches performed at K2K~\cite{Aliu:2004sq}
 and MINOS~\cite{Adamson:2011ig}, and the latest $\nu_\mu \to \nu_e$ appearance 
results from MINOS~\cite{Abe:2011sj}  and T2K~\cite{Adamson:2011qu}.  
This dataset  is insensitive to $\theta_{14}$ and  delimits the vertical band in the left panel of Fig.~1.
To understand this point one should observe that $\nu_4$-driven $\nu_\mu \to \nu_e$  appearance
effects are proportional to the mixing matrix element $U_{\mu 4}$, which is set to zero in our analysis.%

The superposition (left panel of Fig.~1) of the two datasets
sensitive respectively to $\Delta  m^2_{sol}$ and $\Delta  m^2_{atm}$ clearly 
evidences their complementarity in constraining the two mixing angles.
Their synergy manifests quantitatively in their combination displayed in the right panel of Fig.~1.
This provides the strong upper bound
\begin{equation} 
\sin^2 \theta_{14} < 0.04~~~~ (90\%\, \mathrm {C.L.})\,,
\label{eq:bound}
\end{equation}
which constitutes the main result of this report.
For the sake of completeness, we mention that if we had included 
the total rate information from the reactor experiments we would have
obtained a slightly weaker upper bound. For example, adopting the new (higher)
fluxes' estimates the limit would become  $\sin^2 \theta_{14} < 0.05$ at 90\% C.L.
In any case, the anti-correlation existing among the two mixing angles, characteristic
of the solar sector, combined with the independent preference for non-zero $\theta_{13}$,
leads to a strong upper bound on $\theta_{14}$, also destroying any 
weak preference for a non-zero value of this parameter.  As an additional
check of the robustness of the bound in Eq.~(\ref{eq:bound}) we have verified that it is
practically insensitive to the particular choice of the solar  model
used for the calculations. This is important in light of  the yet unresolved 
``metallicity issue'' and its connection with solar neutrino fluxes estimates (see~\cite{Serenelli:2011ea}
for an updated discussion of the topic).

We observe that the bound in Eq.~(\ref{eq:bound}) is not incompatible with the estimates arising from 
the reactor~\cite{Mention:2011rk} and gallium calibration~\cite{Abdurashitov:2005tb,Giunti:2010zu} anomalies.
Rather, lying near their combined best fit~\cite{Mention:2011rk},
it tends to select the lower part of the interval identified by such data.
Probing such relatively low values of $\theta_{14}$ with good precision should be the goal of any 
well-conceived experiment devoted to sterile oscillation searches. 

Finally, we note that our limit is competitive with that one recently established in~\cite{Conrad:2011ce}
using KARMEN and LSND $\nu_e$-carbon cross sections, presenting the additional advantage of being 
independent of the new mass-squared splitting.  This is a unique feature of the solar and  reactor setups
herein considered, where the new oscillations get completely averaged.

\section*{Acknowledgments}
We thank E. Lisi for useful discussions and A. Marrone for  assistance
with the LBL accelerator data analysis. Our work  is supported by the DFG
Cluster of Excellence on  the ``Origin and Structure of the Universe''.

\bibliographystyle{h-physrev4}

\begin{thebibliography}{99}


\bibitem{Palazzo:2011rj}
  A.~Palazzo,
 Phys.\ Rev.\  D {\bf 83}, 113013 (2011)  [arXiv:1105.1705 [hep-ph]].




\bibitem{Giunti:2009xz}
  C.~Giunti and Y.~F.~Li,
  Phys.\ Rev.\  D {\bf 80}, 113007 (2009)
  [arXiv:0910.5856 [hep-ph]].

 
  
\bibitem{Abe:2011sj}
  K.~Abe {\it et al.}  [T2K Collaboration],
  Phys.\ Rev.\ Lett.\  {\bf 107}, 041801 (2011)
  [arXiv:1106.2822 [hep-ex]].

\bibitem{Adamson:2011qu} 
  P.~Adamson {\it et al.}  [MINOS Collaboration],
  Phys.\ Rev.\ Lett.\  {\bf 107}, 181802 (2011)
  [arXiv:1108.0015 [hep-ex]].


\bibitem{Abe:2011fz} 
  Y.~Abe {\it et al.}  [DOUBLE-CHOOZ Collaboration],
  arXiv:1112.6353 [hep-ex].


\bibitem{Fogli:2008jx}
  G.~L.~Fogli, E.~Lisi, A.~Marrone, A.~Palazzo and A.~M.~Rotunno,
  Phys.\ Rev.\ Lett.\  {\bf 101}, 141801 (2008)
  [arXiv:0806.2649 [hep-ph]].
 
\bibitem{Schwetz:2008er}
  T.~Schwetz, M.~A.~Tortola, J.~W.~F.~Valle,
  New J.\ Phys.\  {\bf 10}, 113011 (2008)
  [arXiv:0808.2016 [hep-ph]].
 
\bibitem{GonzalezGarcia:2010er}
  M.~C.~Gonzalez-Garcia, M.~Maltoni, J.~Salvado,
  JHEP {\bf 1004}, 056 (2010)
  [arXiv:1001.4524 [hep-ph]].
 


 
\bibitem{Fogli:2011qn} 
  G.~L.~Fogli, E.~Lisi, A.~Marrone, A.~Palazzo and A.~M.~Rotunno,
  Phys.\ Rev.\ D {\bf 84}, 053007 (2011)
  [arXiv:1106.6028 [hep-ph]].

\bibitem{Schwetz:2011zk} 
  T.~Schwetz, M.~Tortola and J.~W.~F.~Valle,
  New J.\ Phys.\  {\bf 13}, 109401 (2011)
  [arXiv:1108.1376 [hep-ph]].

\bibitem{Machado:2011ar}
  P.~A.~N.~Machado, H.~Minakata, H.~Nunokawa and R.~Z.~Funchal,
  arXiv:1111.3330 [hep-ph].




\bibitem{Formaggio:2011jt}
  J.~A.~Formaggio, E.~Figueroa-Feliciano, A.~J.~Anderson,
   [arXiv:1107.3512 [hep-ph]].

\bibitem{Cribier:2011fv} 
  M.~Cribier, M.~Fechner, T.~Lasserre, A.~Letourneau, D.~Lhuillier, G.~Mention, D.~Franco and V.~Kornoukhov {\it et al.},
  Phys.\ Rev.\ Lett.\  {\bf 107}, 201801 (2011)
  [arXiv:1107.2335 [hep-ex]].

\bibitem{Dwyer:2011xs} 
  D.~A.~Dwyer, K.~M.~Heeger, B.~R.~Littlejohn and P.~Vogel,
  arXiv:1109.6036 [hep-ex].

\bibitem{Novikov:2011gp} 
  Y.~N.~Novikov, T.~Enqvist, A.~N.~Erykalov, F.~v.~Feilitzsch, J.~Hissa, K.~Loo, D.~A.~Nesterenko and L.~Oberauer {\it et al.},
  arXiv:1110.2983 [physics.ins-det].

\bibitem{Egorov}
V. Egorov, talk given at TAUP 2011, International Conference on Topics
in Astroparticle and Underground Physics (Munich, Germany, 2011).    

\bibitem{Gorbachev}
V.V. Gorbachev, talk given at TAUP 2011  (see~\cite{Egorov}).

\bibitem{Ianni}
A. Ianni, talk given at TAUP 2011 (see~\cite{Egorov}).

\bibitem{Lasserre}
T. Lasserre, talk given at TAUP 2011 (see~\cite{Egorov}).


\bibitem{Bowden}
N. Bowden, talk given at SNAC 2011, Workshop on Sterile neutrinos at the crossroads (Blacksburg, Virginia, USA 2011).



%

\bibitem{Mueller:2011nm} 
  T.~A.~Mueller, D.~Lhuillier, M.~Fallot, A.~Letourneau, S.~Cormon, M.~Fechner, L.~Giot and T.~Lasserre {\it et al.},
  Phys.\ Rev.\ C {\bf 83}, 054615 (2011)
  [arXiv:1101.2663 [hep-ex]].


\bibitem{Huber:2011wv}
  P.~Huber,
  Phys.\ Rev.\  {\bf C84}, 024617 (2011)
  [arXiv:1106.0687 [hep-ph]].


\bibitem{deGouvea:2008qk} 
  A.~de Gouvea and T.~Wytock,
  Phys.\ Rev.\ D {\bf 79}, 073005 (2009)
  [arXiv:0809.5076 [hep-ph]].


\bibitem{Apollonio:2002gd}
M.~Apollonio {\em et~al.} (CHOOZ Collaboration),
\newblock Eur. Phys. J. C {\bf 27}, 331 (2003).


\bibitem{reactor}
F.~Ardellier {\em et~al.} (Double Chooz Collaboration),
\newblock hep-ex/0606025; M.~C. Chu (Daya Bay Collaboration), 
arXiv:0810.0807;  Y.~Oh (RENO Collaboration),
Nucl.\ Phys.\ Proc.\ Suppl.\  {\bf 188} (2009) 109.



\bibitem{Dydak:1983zq}
  F.~Dydak {\it et al.},
  Phys.\ Lett.\  B {\bf 134}, 281 (1984).


\bibitem{Mahn:2011ea}
 K.~B.~M.~Mahn {\it et al.}  [SciBooNE and MiniBooNE Collaboration],
  arXiv:1106.5685 [hep-ex].
  
\bibitem{Maltoni:2007zf} 
  M.~Maltoni and T.~Schwetz,
  Phys.\ Rev.\ D {\bf 76}, 093005 (2007)
  [arXiv:0705.0107 [hep-ph]].

\bibitem{Adamson:2011ku} 
  P.~Adamson {\it et al.}  [MINOS Collaboration],
  Phys.\ Rev.\ Lett.\  {\bf 107}, 011802 (2011)
  [arXiv:1104.3922 [hep-ex]].



  
  





 
\bibitem{Schwetz:2011qt}
  T.~Schwetz, M.~Tortola, J.~W.~F.~Valle,
  New J.\ Phys.\  {\bf 13}, 063004 (2011)
  [arXiv:1103.0734 [hep-ph]].
 


  
\bibitem{Balantekin:2008zm}
  A.~B.~Balantekin, D.~Yilmaz,
  J.\ Phys.\ G {\bf G35}, 075007 (2008)
  [arXiv:0804.3345 [hep-ph]].



\bibitem{Aliu:2004sq}
  E.~Aliu {\it et al.} [ K2K Collaboration ],
  Phys.\ Rev.\ Lett.\  {\bf 94}, 081802 (2005)
  [hep-ex/0411038].

\bibitem{Adamson:2011ig}
  P.~Adamson {\it et al.} [ The MINOS Collaboration ],
  Phys.\ Rev.\ Lett.\  {\bf 106}, 181801 (2011)
  [arXiv:1103.0340 [hep-ex]].


\bibitem{Serenelli:2011ea}
  A.~Serenelli,
   [arXiv:1109.2602 [astro-ph.SR]].
 



\bibitem{Mention:2011rk} 
  G.~Mention {\it et al.},
  Phys.\ Rev.\ D {\bf 83}, 073006 (2011)
  [arXiv:1101.2755 [hep-ex]].


 
\bibitem{Abdurashitov:2005tb}
  J.~N.~Abdurashitov {\it et al.},
  Phys.\ Rev.\  C {\bf 73}, 045805 (2006)
  [arXiv:nucl-ex/0512041].

\bibitem{Giunti:2010zu} 
  C.~Giunti and M.~Laveder,
  Phys.\ Rev.\ C {\bf 83}, 065504 (2011)
  [arXiv:1006.3244 [hep-ph]].





\bibitem{Conrad:2011ce}
  J.~M.~Conrad, M.~H.~Shaevitz,
  [arXiv:1106.5552 [hep-ex]].

  





\end{thebibliography}

\end{document}